# Transport barrier formation by LHCD on TRIAM-1M

K.HANADA[1)], A.IYOMASA[1)], H.ZUSHI[1)], M.HASEGAWA[1)], K.SASAKI[2)], H.HOSHIKA[2)], K.NAKAMURA[1)], M.SAKAMOTO[1)], K.N.SATO[1)], H.IDEI[1)], S.KAWASAKI[1)], H.NAKASHIMA[1)], and A.HIGASHIJIMA[1)]

1) Advanced Fusion Research Center, Research Institute for Applied Mechanics, Kyushu University, Kasuga 816-8580, Fukuoka, Japan
2) Interdisciplinary Graduate School of Engineering Sciences, Kyushu University

E-mail:hanada@triam.kyushu-u.ac.jp

**Abstract**: Internal transport barrier (ITB) has been obtained in full lower hybrid current driven (LHCD) plasmas on a superconducting tokamak, TRIMA-1M (R=0.84m, a x b=0.12mx0.18m, $B_T$<8T). The formation of ITB depends on the current density profile, j(r), varied by the power deposition of the lower hybrid (LH). The plasma with ITB can be maintained by the LH power deposited around the foot point of ITB up to 25 sec, which corresponds to more than 100 times of current diffusion time, $\tau_{L/R}$. ITB is terminated by the reduction of current drive efficiency caused by metal impurities accumulation. In some condition, self-organized slow sawtooth oscillations (SSSO) of plasma current, density, temperature, and so on with the period comparable to the current diffusion time have been also observed during ITB discharge. The oscillation has the capability of particle exhaust, as the result, it may play an role in the avoidance of the impurity accumulation and the dilution in the future steady state fusion plasma with ITB, as the edge-localized mode in H-mode.

## 1. Introduction

Steady state operation of high performance plasma is one of the key issues to realize cost-effective fusion power plant. High performance plasma with the transport barrier is a good candidate from the view of plasma confinement [1-7]. Especially it is important that the high performance plasma was maintained by fully non-inductive current drive. Some experimental observations have been shown to achieve the good performance plasma for the longer duration than the current diffusion time [8, 9]. Many studies of internal transport barrier (ITB) were carried out in many devices and it was found that ITB is common phenomena [10-13]. In tokamak plasmas, transport coefficients have strong relation to current density profile, that is magnetic shear [4, 5], and the current profile control is important tool for the control the performance of the plasma. Especially lower hybrid current drive (LHCD) is powerful tool for the control of current profile and in fact some experiments show the promising possibility [14, 15]. The avoidance of impurities accumulation is one of key issue to execute steady state operation of ITB [16, 17]

Recently, self-organized sinusoidal oscillation could be observed in long duration plasma maintained by fully non-inductive LHCD on Tore Supra [18]. This oscillation could be understood as a predator-pray instabilities derived from the variation of the transport coefficient related to current density profile as the following. The power deposition of LHCD depends on the electron temperature and the electron temperature is also controlled by the transport coefficient related to current density profile. The relation was formed a predator-prey relation and as the result the oscillation in the electron temperature and current density profile appears in core region of the non-inductive long duration discharge. The growth rate of the oscillation is zero and there are no effects in the exhaust of the accumulated impurities and it may not play an essential role in the particle exhaust of the plasma with ITB [18].

On TRIAM-1M [19], the formation of ITB and the oscillation related with ITB formation has been obtained in ECD plasmas [20-22]. The plasma with ITB can be maintained by the LH power deposited around the foot point of ITB up to 25 sec, which corresponds to more than 100 times of current diffusion time, $\tau_{L/R}$. The oscillation has the capability for the exhaust of the particles including the impurities. In fact, a transition to high performance plasma with ITB could be achieved under the high influx of impurities.

In section 2, the brief introduction of the experimental apparatus is shown and the experimental results concerning the formation of ITB is described in section 3 and consequently the results of the oscillation related ITB formation is shown in the section 4. The summary is described in the section 5.

## 2. Experimental Apparatus

TRIAM-1M (R=0.84m, a x b=0.12m x 0.18m) is a small size high magnetic field tokamak with 16 superconducting toroidal magnetic field coils made of $Nb_3Sn$, which produces the toroidal magnetic field, $B_T$, up to 8T in steady state. Experiments described in this paper are carried out by two 8.2 GHz LHCD systems. One system (System_1) has the capability to change the power during the discharge and change the refractive index along the magnetic field, $N_{//}$, of LHW shot by shot. Another system (System_2) has the capability to control the value of $N_{//}$ of LHW during a discharge with 10 degree per a second as well as the power.

Main diagnostics in this paper are a soft X-ray (SXR) detector array composed of 19ch silicon surface barrier (SSB) diodes sensitive to photon in the range of 0.07keV-30keV and a hard X-ray (HXR) detector array composed of 7ch NaI scintillators sensitive to photon with 40-500keV. Low energy part of HXR is absorbed by the vacuum window made of Aluminum. The NaI scintillators have been arranged along the major radius. The viewing chord sets to the perpendicular direction to the toroidal magnetic field

## 3. The formation of internal transport barrier

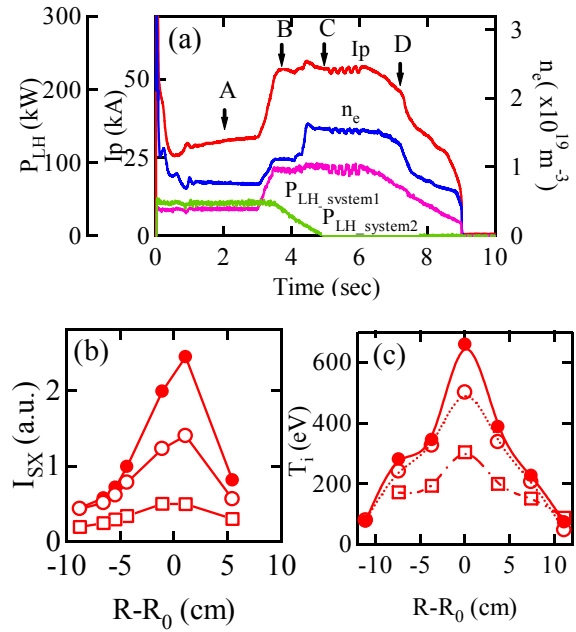

Typical waveforms of the discharge with ITB are shown in Fig.1(a). The time slices of the profile of soft X-Ray and ion temperature, $T_i$, are also shown in Figs.1(b) and (c). The plasma was maintained by the combination of two LH waves with the different $N_{//}$ controlled by the phase difference of the adjacent waveguide at the launcher. All of the plasma current was driven by these LH power non-inductively at any time except the start-up phase of the discharge. The LH power of System_1 was gradually raised from 3 s to 3.5s by 50kW. As the result, both plasma current and electron density follow the increase of injected LH power. In this duration, the current drive efficiency did not change with that before the increment of the LH power of System_1. From 3.5s, the LH power of the System_2 is gradually reduced and the abrupt increase of the electron density can be observed around 4.2s. At that time, the profile of SXR Intensity, $I_{SX}$, is also peaked as shown in Fig. 1(b) and the current drive

Fig.1 (a): Time evolution of plasma current (red line), electron density (Blue line), LH power of System_1 (pink line) LH power of System_2 (green line) are plotted in the upper figure. (b): The profiles of soft X-ray intensity at the time marked as A (open squares), B (open circles) and C (closed circles) in the upper figure are shown in the lower figure. (c) The profiles of ion temperature at the time marked as A (open squares), C (open circles) and D (closed circles) in the upper figure are shown.

efficiency is also improved. When the LH power of System_2 is not reduced, the peaked $I_{SX}$ profile is not obtained. While the LH power of System_1 is not raised, the peaked $I_{SX}$ profile is not also obtained. These results show that the LH power of System_1 assists the formation of the peaked $I_{SX}$ profile and that of System_2 obstructs it. The peaked $I_{SX}$ profile shows that the electron pressure is peaked as shown in the rapid increase of electron density. When the peaked $I_{SX}$ profile is formed, the total of the injected LH power to the plasma is reduced. This shows that the peaked electron pressure is not made by the increment of the deposited power to the center of the plasma, but by the reduction of transport coefficient around r=5 cm, where is around the foot of ITB in the plasma. The ion temperature, $T_i$, at the center of the plasma also increases as shown in Fig. 1(c) at the time of C in Fig. 1(a). The ion transport is tentatively improved during more during the reduction of the power of LH as shown by closed circles in Fig. 1(c). The high $T_i$ state is called as "Meta stable" and it can be maintained by the control of the power level and $N_{//}$ of the injected LH wave. After the transition, the LH power of System_1 is gradually reduced and the back-transition takes place around 7 s. After the back-transition, $n_e$, $I_P$ and $T_i$ return to the previous levels and finally the further reduction of the LH power makes the termination of the plasma. The LH power required to maintain the plasma current is about 30kW.

To investigate the difference of the effect of each LH power to the plasma, the modulation of the LH power during the discharge was carried out. The LH power modulation was executed from 5s to 6s as shown in Fig. 1. The LH power was modulated every 100ms by 15% of the power. The plasma current and electron density were also modulated as shown in Fig.1. The intensities of HXR, $I_{HX}$, correspond to the signal of the pressure of tail electrons around 40keV. It is noted that $I_{HX}$ is also proportional to the bulk electron density and $Z_{eff}$. To remove the effect of electron density and $Z_{eff}$ profile to $I_{HX}$ signal, the inversed decay time, $1/I_{HX} dI_{HX}/dt$ is calculated and they are plotted as the function of minor radius as shown in Fig.2. When the inversed decay time is multiplied by energy stored of tail electrons, the power deposition of LHW will be calculated. The power deposition of System_1 is mainly deposited at the position of r=5cm in any cases, it suggests that the LH power of System_1 is mainly deposited at the position of r=5cm and it plays a role in making broad current profile. While the LH power of System_2 is deposited at the center of plasma and it makes peaked current profile. This suggests that the formation of ITB is assisted by the LH current drive at the foot point of ITB.

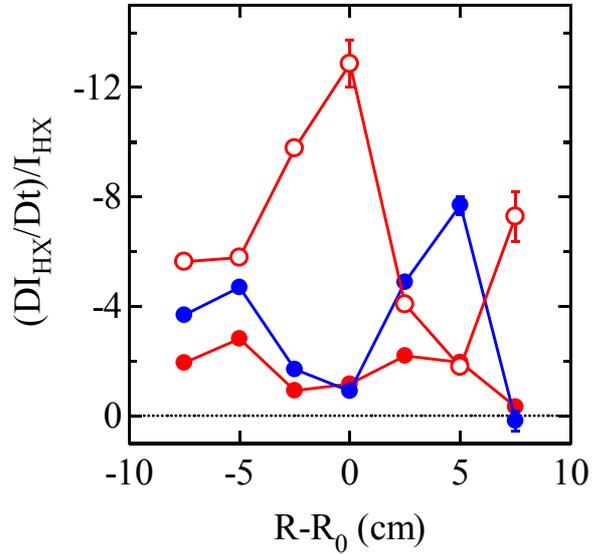

Fig.2 Decay time profile of System_1 with ITB (closed blue circles), System_1 without ITB (open blue circles), and System_2 without ITB (open red circles) during the LH power modulation experiment are plotted as the function of minor radius.

### 4. Maintenance of ITB

The long duration discharge with ITB by use of only System_1 can be achieved up to 25 s, which corresponds to 100 times of current diffusion time, $\tau_{L/R}$ is shown in Fig. 3. A back-transition takes place during the discharge, and the discharge can be maintained by LHCD for 36 sec. The influx of ferrite estimated by the line intensity of FeI increases during the discharge gradually. In the other discharge, just before the termination of ITB, a number of grains of Molybdenum put into the plasma and plasma performance become to be worse. The grains come from the poloidal limiter made of Molybdenum. The time evolution of the line intensity of Mo XIII inversely synchronizes with that of the plasma current. These indicate that the metal impurity accumulation plays an essential role in the back-transition from high performance plasma with ITB.

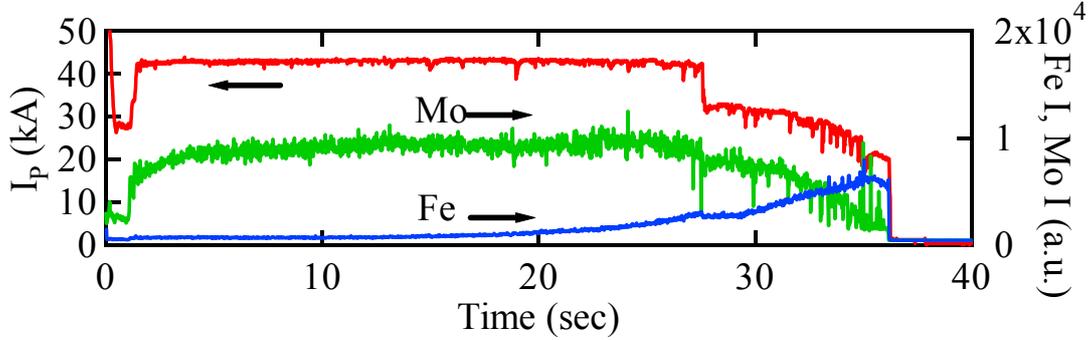

Fig. 3 Time evolutions of Ip, Fe I, and Mo I of the long duration ITB discharge are shown. On the middle of the discharge, ITB was terminated at 27 sec and then the plasma was terminated at 36 sec from the start of discharge.

## 5. Self-organized Oscillation related with ITB

The new kind of self-organized slow sawtooth oscillation (SSSO) has been observed in long-duration fully non-inductive LHCD plasmas. SSSO has been obtained only in high performance plasma with ITB. The period of the oscillation is comparable to the current diffusion time, $\tau_{L/R}$, of the plasma (a few handreds ms). The crashes of SSSO is sometimes accompanied by precursor, which is similar to the predator-pray instability observed in Tore Supra. This precursor has no helical structure and the period is similar to the change of the current profile (a few tens ms).

The typical waveforms of the plasma parameters are shown in Fig. 4. The plasma current is maintained by lower hybrid wave (LHW) of the peak $N_{//}$=1.8 excited by System_1, where $N_{//}$ is the refractive index along the magnetic field line. From the beginning of plasma, the additional LHW excited by another LHCD device of the frequency in 8.2GHz is injected into the plasma. The value of $N_{//}$ of System_2 is shifted from 1.8 (0-2sec) to 2.0 (4sec-the end of the discharge) during the discharge. The net injection power of RF increases at 1.2 sec because of the improvement of the coupling between microwave and the plasma, and as the result the first jump of $I_P$ and $n_e$ occurs around 1.2 sec as shown in Figs. 4(a) and (b), where the relative intensity between SXR emitted at the center chord and at the peripheral chord does not change so much as shown in Fig. 4(d). The second jump of $I_P$ and $n_e$ is observed around 1.5sec, although the net LH power does not change [20-22]. At the second jump, the intensity ($I_{SX}$) of SXR at the center chord burgeons and that at the peripheral chord decreases slightly. This shows that the formation of steep gradient of $n_e$ and electron temperature, $T_e$, occurs at the time of the second jump, which shows the formation of ITB in the plasma. The position of ITB foot is estimated by the profile of SXR and it locates around r=6 cm (r/a=0.5), where r and a are the distance between line of sight and the plasma center and minor radius of the plasma, respectively. The intensity ($I_{HX}$) of hard X-ray (HXR) at the center of the plasma also burgeons as shown in Fig. 3(d). This suggests that ITB of $n_e$ is formed, rather than $T_e$. The formation of ITB has the strong relation to the LH power [20] and $\tilde{N}_{//}$ [21]. With the reduction of $N_{//}$, the deposition of LHW is far from the center of plasma and the current drive efficiency, $\eta_{CD}$, degrades. When the value of $N_{//}$ of the additional LHW sets to more than 2.4, ITB is not able to be achieved. It is found that the power deposited at the center of the plasma prevents from forming ITB, where the power deposition can be measured with the variation of HXR during the power modulation. This shows that the formation of ITB depends on the current profile driven by the injected LHW. The RF power of $N_{//}$=1.8 mainly deposits at the ITB foot (r/a=0.5).

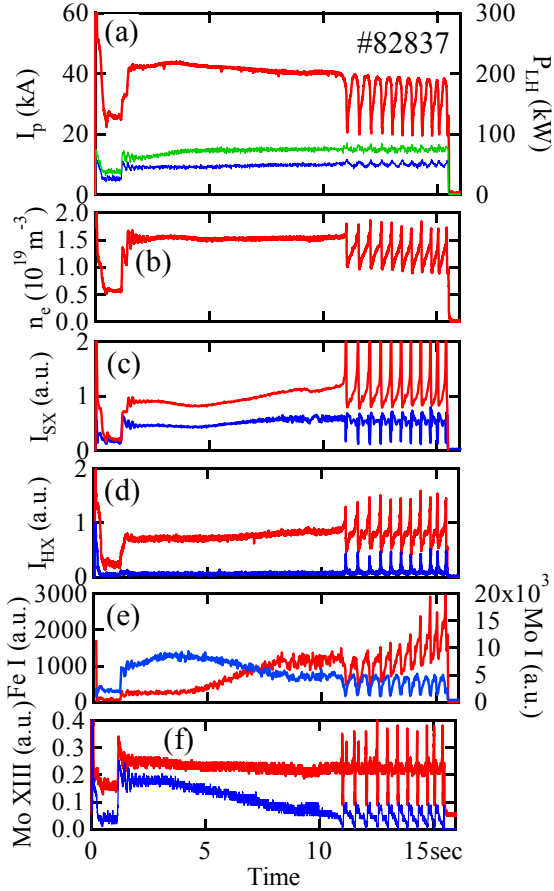

Fig. 4 The typical waveforms of (a) plasma current, IP and the net injected LH power, PLH, (b) electron density, ne, (c) Intensity of SXR at r=0cm (red line) and 7.8cm (blue line), (d) Intensity of HXR at r=0cm (red line) and 5cm (blue line), (e) Photon count of FeI (red line) and MoI (blue line), and (f) VUV signal of MoXIII at r=2.5cm (red line) and 5cm (blue line).

The plasma current is reduced gradually, because the reduction of $\eta_{CD}$ occurs due to the shift of the $N_{//}$ from 2 to 4 sec. Since 4 sec, the reduction of $\eta_{CD}$ seems to be derived from the increase of $Z_{eff}$. The intensity of FeI ($I_{FeI}$), which is proportional to the influx of ferrite, increases gradually as shown in Fig. 3(e) and simultaneously the intensity of MoI ($I_{MoI}$) is reduced, where the values of $I_{FeI}$ and $I_{MoI}$ is measured with a spectroscopy viewing the movable limiter from the bottom port of #1. Ferrite come form the movable limiter installed on the upper part of #1 port. Energetic electrons produced by LHW are directly attacked to the movable limiter. The movable limiter is made of SS covered by Molybdenum. The energy and orbit of the energetic electrons depends on the value of $\tilde{N_{//}}$. The waveform of the intensity of Mo XIII ($I_{MoXIII}$) outside of the ITB region is reduced and it is similar to $I_{MoI}$. However $I_{MoXIII}$ around the ITB foot does not decrease as shown in Fig. 3(f). This indicates that the metal impurities accumulate in the core region as reported in the other devices [10, 11]. Unfortunately we do not measure the radiation of the ferric ion. However ferric ion will accumulate more inside ITB because of the increase of the influx of ferrite as shown in Fig. 4(e). Surely the intensity of SXR at the center of the plasma suggests that the value of $Z_{eff}$ gradually increases.

Around 11 sec, an oscillation appears in plasma current, density and so on as sown in Fig. 4. The periods of the oscillation is in the range of a few hundreds ms and it corresponds to the current diffusion time (~200ms) of the plasma. This oscillation occurs in the self-organized manner and the oscillation is sometimes accompanied by the periodic predator-pray type instabilities observed in Tore Supra [18] just before the crash. The termination of the discharge is not due to the long-duration oscillation, but to the shut-down of the LH power.

The region of the appearance of the self-organized oscillation can be shown on the map of $I_{FeI}$ and the value of $N_{//}$ of the additional LHW. The LHW with low $N_{//}$ has the preferable capability to the current drive and its current drive efficiency is better than that of the LHW with high $N_{//}$. The plasma with ITB can be achieved in the plasma with the additional LHW with $N_{//}$<2.4. When the value of $N_{//}$ of the additional LHW is less than 2.4, it is difficult to make the plasma with ITB. When the value of influx of ferrite is larger than a certain value, it is difficult to maintain the plasma with ITB. However, SSSO can work on the assistance with the formation of ITB even in the high influx of ferrite as shown in Fig. 6. This indicates that

SSSO plays an essential role in the exhaust of the metal impurities from the inside of ITB. This action of SSSO may work on the avoidance of the impurity accumulation and the dilution in steady state operation of fusion plasma with ITB.

## 6. Summary

ITB has been obtained in full LHCD plasmas on a superconducting tokamak, TRIMA-1M (R=0.84m, a x b=0.12m x 0.18m, $B_T$<8T). The plasma with ITB can be maintained by the LH power deposited around the foot point of ITB up to 25 sec, which corresponds to more than 100 times of current diffusion time, $\tau_{L/R}$. The metal impurity accumulation prevents us from maintaining the discharge with ITB for long time. The LH power deposited around ITB foot assists the formation of ITB and the LH power deposited at the center of the plasma obstacles it. This suggests that the hollow current profile is suitable to form ITB. Self-organized slow sawtooth oscillations (SSSO) of plasma current, density, temperature, and so on with the period comparable to the current diffusion time have been also observed in long duration full LHCD discharges. The oscillation appears only in the high performance plasma with ITB. The action of particle exhaust caused by SSSO is preferable to avoid the impurity accumulation.

## 7. Acknolegement

This work has been partially performed under the framework of joint-use research in RIAM Kyushu University and the bi-directional collaboration organized by NIFS.
This work is partially supported by a Grant-in-Aid for Scientific Research from Ministry of Education, Science and Culture of Japan.